\documentclass[amsmath,amssymb,aps,floatfix,prl,superscriptaddress,preprint]{revtex4-2}

\usepackage[utf8]{inputenc} 
\usepackage[T1]{fontenc} 
\usepackage{amsfonts} 
\usepackage{booktabs} 
\usepackage{hyperref} 
\usepackage{url} 
\usepackage{microtype} 
\usepackage{nicefrac} 
\usepackage{paralist} 
\usepackage{xcolor}
\usepackage{graphicx}
\usepackage{float}
\usepackage{wrapfig}
\usepackage{ulem}
\usepackage{fullpage}

\usepackage{lmodern}
\usepackage{ulem}

\begin{document}
	
	\title{Unsupervised topological learning approach of crystal nucleation}
	
	\author{S\'ebastien Becker }
	\affiliation{Université Grenoble Alpes, CNRS, Grenoble INP, SIMaP\\
		F-38000 Grenoble, France}
	\affiliation{Université Grenoble Alpes, CNRS, Grenoble INP, LIG\\
		F-38000 Grenoble, France}
	\author{Emilie Devijver}
	\affiliation{Université Grenoble Alpes, CNRS, Grenoble INP, LIG\\
		F-38000 Grenoble, France}
	\author{R\'emi Molinier}
	\affiliation{Université Grenoble Alpes, CNRS, IF\\
		F-38000 Grenoble, France}
	\author{No\"{e}l Jakse}
	\affiliation{Université Grenoble Alpes, CNRS, Grenoble INP, SIMaP\\
		F-38000 Grenoble, France}
	
	\maketitle
	\section*{ }
	\textbf{
		Nucleation phenomena commonly observed in our every day life are of fundamental, technological and societal importance in many areas, but some of their most intimate mechanisms remain however to be unravelled. Crystal nucleation, the early stages where the liquid-to-solid transition occurs upon undercooling, initiates at the atomic level on nanometre length and sub-picoseconds time scales and involves complex multidimensional mechanisms with local symmetry breaking  that can hardly be observed experimentally in the very details. To reveal their structural features in simulations without \textit{a priori}, an unsupervised learning approach founded on topological descriptors loaned from persistent homology concepts is proposed. Applied here to monatomic metals, it shows that both translational and orientational ordering always come into play simultaneously when homogeneous nucleation starts in regions with low five-fold symmetry. It also reveals the specificity of the nucleation pathways depending on the element considered, with features beyond the hypothesis of Classical Nucleation Theory. 
				}

	Understanding homogeneous crystal nucleation under deep undercooling conditions remains a formidable issue, as crystallization is essentially heterogeneous in nature and initiated from impurities, surfaces, or near grain boundaries that often hinder its occurrence \cite{Sosso2016,Kelton2010}. Unreachable until very recently, experimental observations of early stages of nuclei was achieved by a \textit{tour de force} using time tracking of three-dimensional (3D) Atomic Electron Tomography \cite{Zhou2019} of metallic nanoparticles. Those complex phenomena remain to date out-of-reach experimentally for bulk systems, thus hindering our theoretical understanding. This line of research still belongs mostly to the domain of atomic-level simulations and more particularly to molecular dynamics (MD) with generic interaction models \cite{Auer2001,ten1995}. To reach statistically meaningful events, large scale simulations are required. This still remains challenging as only few studies are providing now million-to billion-atom simulations for monatomic metals \cite{Sosso2016}. 

	To identify translational and orientational orderings during homogeneous nucleation in MD simulations, an unsupervised learning approach \cite{Ceriotti2019} based on topological data analysis (TDA) signatures was developed through persistent homology (PH) \cite{Carriere2015, Motta2018}. PH is an intrinsically flexible, yet highly informative, tool which detects meaningful topological features deduced from atomic configurations. It was successfully applied very recently to characterise structural environments in metallic glasses \cite{Hirata2020}, ice \cite{Hong2019} and complex molecular liquids \cite{Sasaki2018}. Always used as a structural analysis in these studies, the originality here is to use PH as a translational and rotational invariant descriptor to encode the local structures required for the clustering method. For the latter a model-based method is used, namely Gaussian Mixture Models (GMM) \cite[Chapter 14]{tibshirani} (already used with success to analyse MD simulations \cite{Boattini2020}) and its estimation by an Expectation Maximization (EM) algorithm \cite{Dempster}. The number of clusters \cite{warningCluster} is selected by Integrated Criterion Likelihood (ICL, \cite{ICL}), a refinement for clustering of Bayesian Integrated Likelihood (BIC, \cite{BIC}). The inferred model from the method called hereafter TDA-GMM, is used to identify and describe the structural and morphological properties of the nuclei as well as their liquid environment at various steps of the crystal nucleation. 
	
	\begin{figure}[tb!]
		\includegraphics[scale=0.7]{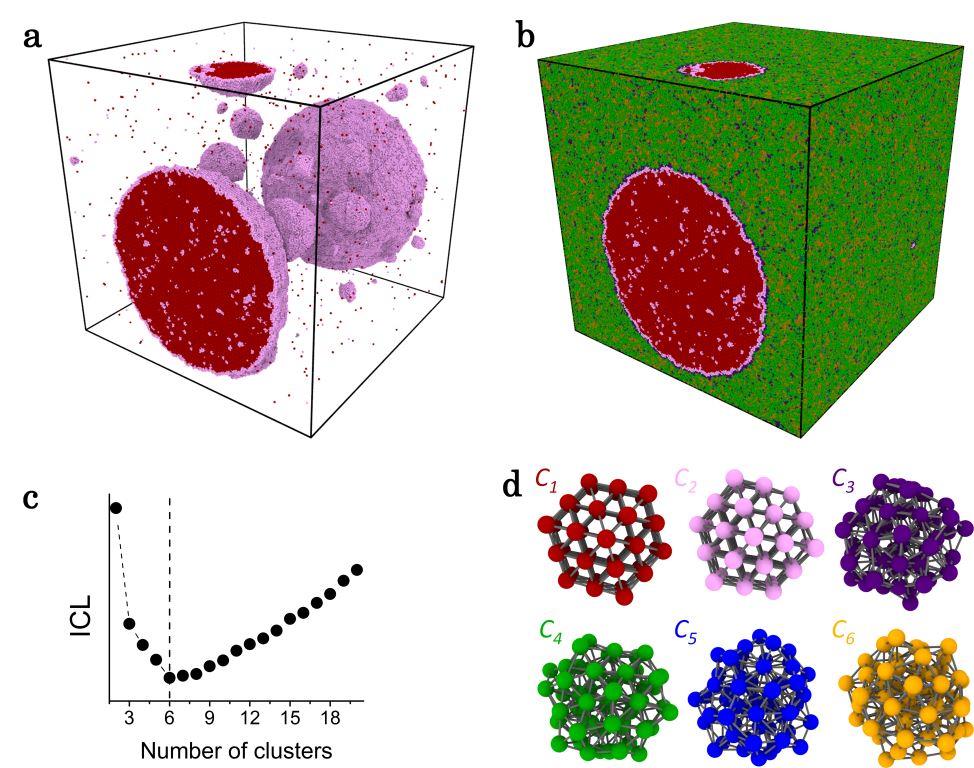}
		\caption{\textbf{Unsupervised learning of homogeneous nucleation.} Snapshot of a ten-million atom MD simulation of Ta during nucleation along the $T=1900$~K isotherm (a and b). Independent local atomic structures within a cut-off-radius of $6.8$ \AA{} form a train set represented in the descriptor space by $215$ PH components up to the second order. (c) Evolution of the ICL criterion as a function of number of clusters is used to get autonomously the optimal number of clusters  shown in (d). In (a) the snapshot is represented only with atoms in cluster $C_1$ and cluster $C_2$ revealing all nuclei (see text), while in (b) atoms of all clusters are displayed showing that those in cluster $C_3$ are located mainly at the border of the nuclei and $C_4$, $C_5$ and $C_6$ correspond to the surrounding liquid with various topological characteristics.}
		\label{fig:Fig1-method}
	\end{figure}
	
	With this unsupervised approach, the homogeneous nucleation process was studied in three monatomic metals chosen for the variety of their underlying crystalline phase, namely body-centered cubic (bcc) for Ta, face centred-cubic (fcc) for Al, and hexagonal-closed packed (hcp) for Mg. Large-scale molecular dynamics simulations \cite{LAMMPS} comprising one and ten million atoms were performed with a similar procedure used in our preceding work on pure Zr \cite{Becker2020} and described in more details in  Methods Section. Figure \ref{fig:Fig1-method} depicts the methodology applied here to Ta. A rapid quenching at constant pressure brings the liquid from $T=3300$ K down to $T=1900$~K close to the time-temperature-transformation (TTT) nose. Crystal nucleation is observed along an isothermal process during which a configuration of the simulation is chosen for the clustering. As it contains many nuclei with different sizes and a substantial amount of liquid, it is considered as representative of the phenomenon. From its inherent structure \cite{Stillinger1982}, a training set of $5$ $000$ non overlapping local spherical structures  within a cutoff radius of $6.8$ \AA{} was sampled for the unsupervised learning (see Supplementary Information), with the constraints of covering the entire simulation box uniformly and randomly. Among all possible sets upon applying the GMM, the one with $6$ clusters shown in \ref{fig:Fig1-method} (d) was automatically inferred to be representative of the system based on the minimum ICL criterion \ref{fig:Fig1-method}(c). The snapshot of the simulation box in Fig. \ref{fig:Fig1-method}(a) displays only atoms of type $C_1$ and $C_2$, as they show clearly a crystalline order, refraining at this stage from characterising it. They reveal all nuclei as it will be seen below, along with their structure, size and morphology out of the simulation box displayed in Fig. \ref{fig:Fig1-method}(b). From this model, each atom of each configuration generated by the MD simulation can be assigned to one of the six clusters (the one with the highest probability). Such a clustering training is performed independently for each metal and shows that more than $99.99$ \% of the structures have a probability to belong to the most probable Gaussian component greater than $0.999$, even for structures not in the initial training set.
	
	\begin{figure}[tb!]
		\includegraphics[scale=0.13]{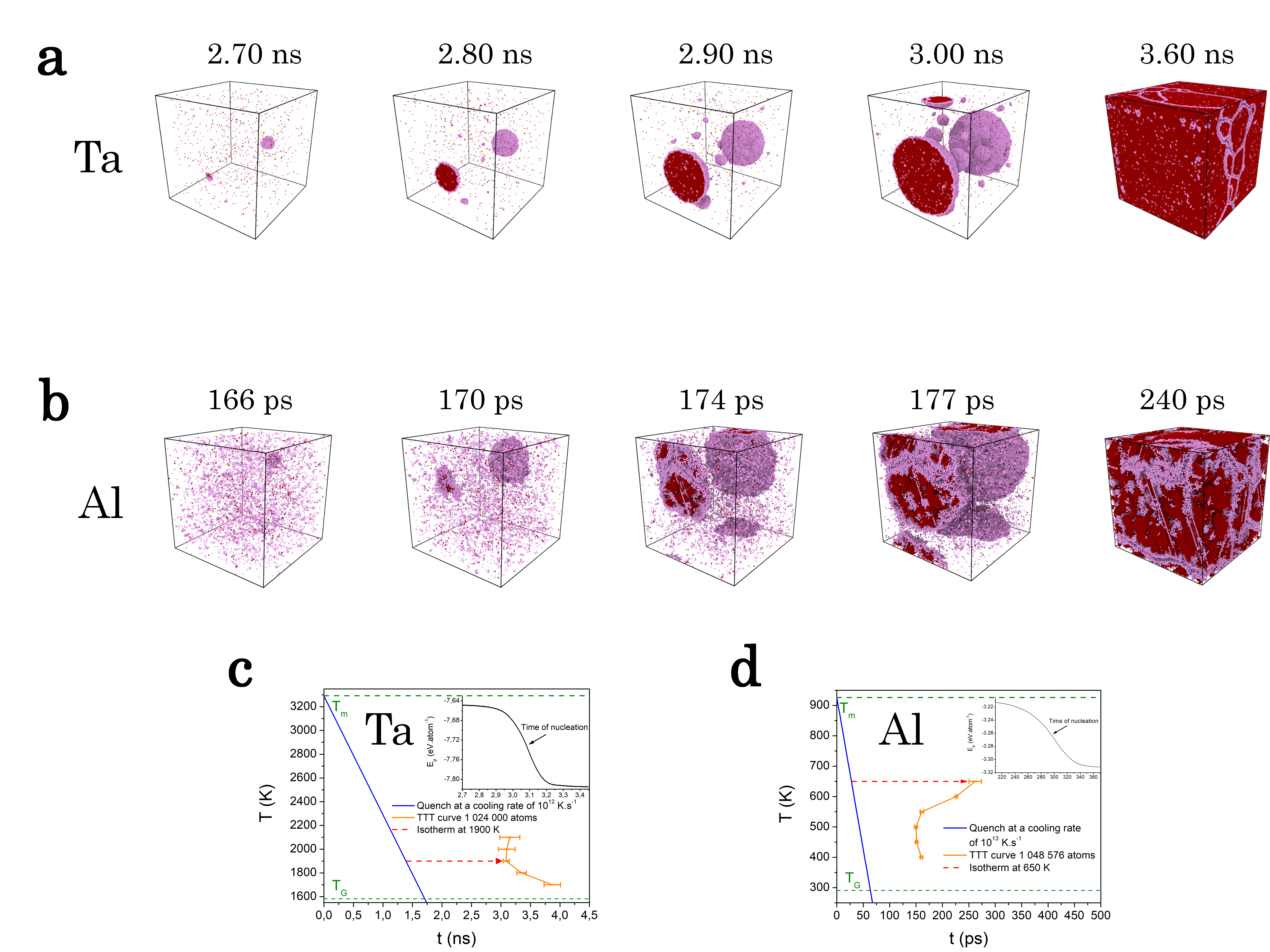} 
		\caption{\textbf{Homogeneous nucleation in Ta and Al undercooled liquids.} Snapshots of the molecular dynamics simulations for Ta (a) and Al (b) with respectively $10$ and $1$ million atoms, during isothermal nucleation at different times for temperatures close to the nose of the Time-Temperature Transformation (TTT) for Ta (c) and Al (d). From stored configurations during fast cooling (blue curves), nucleation events along several isotherms were observed by monitoring the sharp drop of the internal energy (insets in (c) and (d)). The average nucleation times $\tau_N$ (symbols) were determined from $5$ independent simulations for each temperature giving the TTT curves in the vicinity of the nose (orange lines).}
		\label{fig:Fig2-observation}
	\end{figure}
	
	Figure \ref{fig:Fig2-observation} shows typical homogeneous nucleation events in undercooled Ta and Al during an isothermal process close to the nose of the TTT, which can be done by standard MD simulations without the need of an accelerated methods such as the Forward-flux sampling method \cite{Allen2009}. The liquids above the melting point $T_{M}$ were first quenched down at ambient pressure to the glass transition sufficiently rapidly to avoid nucleation (see Table S1 in Supplementary Information). From stored configurations during cooling, the TTT curves in the vicinity of the nose were built from observation of the nucleation along several isotherms as shown in Figs. \ref{fig:Fig2-observation}(c) and (d). An isotherm slightly above the TTT nose is chosen for the analysis, \textit{i.e.} $T=1900$ K for Ta and $T=650$ K for Al. From chosen configurations during the nucleation and growth process, the clustering is obtained from application of the corresponding trained model as described above. For all metals considered here, strongly growing fraction of mainly two clusters, concomitant to the sharp drop of the energy, is observed. For Ta and Al, only local structures belonging to these clusters are drawn in Figures \ref{fig:Fig2-observation}(a) and (b), revealing evidently the nuclei and their evolution in time, recalling that solely the topological vector are describing the local structure. The nuclei morphologies show globular shapes that are rather spherical, characteristic of high $\Delta T$, although obviously not strictly as revealed more quantitatively from a convex hull analysis. Interestingly, atoms from one of the two clusters (coloured in red) are mainly located inside the nuclei while atoms from the second one (coloured in pink) steadily remain essentially at the border upon growing. They stay finally at grain boundaries after full solidification of the simulation boxes. Its appearance inside the nuclei reveals also the presence of defaults, as it will be examined below. 
	
	The simulations of homogeneous nucleation shown in Fig. \ref{fig:Fig2-observation} were performed with $10$ and $1$ million atoms for Ta and Al, respectively. In both cases, the vast majority of the embryos dissolve back to the liquid while those attaining the critical size are rare and grow. The larger simulation box for Ta allows to follow the nucleation process for a longer time, sufficient to observe more secondary nucleation events \cite{Shibuta2017}. Direct estimation of the critical size is still unreachable by experiment, as nuclei can be detected only at larger size \cite{Zhou2019}. This is also scarcely studied by MD simulation as it is not easy to define their boundary from the surrounding liquid \cite{tenWolde1996,Baez1995}, especially in the case of non-spherical or ramified shape \cite{Toxvaerd2020}. Here, the size distribution of nuclei was obtained by counting the number of atoms in overlapping structures identified as red and pink clusters within the cut-off radius. An estimation of the critical size was inferred from the nuclei's size that persists between the first and second configurations shown in Fig. \ref{fig:Fig2-observation}, at least without loosing atoms they contained initially. As it can be seen for Ta on Fig. \ref{fig:Fig1-method}(d), the local structures of the two clusters forming the nuclei are unambiguously crystalline (with only a slight distortion for structures from cluster $C_2$) giving a clear definition of them. This is repeated in the subsequent consecutive pairs of configurations to refine statistics, and the results for all metals are gathered in Table S1 in Supplementary Information. For Ta, embryos with size less than $120$ atoms always dissolve back to the liquid while the few nuclei found with size larger that $150$ atoms always grew. Similar values of the critical radius were determined very recently for bcc Fe and fcc Cu \cite{Louzguine2020} and fcc Zn \cite{Zhou2016} in similar high $\Delta T$ regime. For Al and Mg the simulations were performed at lower $\Delta T$ yielding obviously larger critical nuclei which are consistent with the Lennard-Jones case \cite{ten1995,tenWolde1996} and also with Al but somewhat lower with respect to recent MD simulations \cite{Mahata2018}.  
	
	\begin{figure}[tb!]
		\includegraphics[scale=0.11]{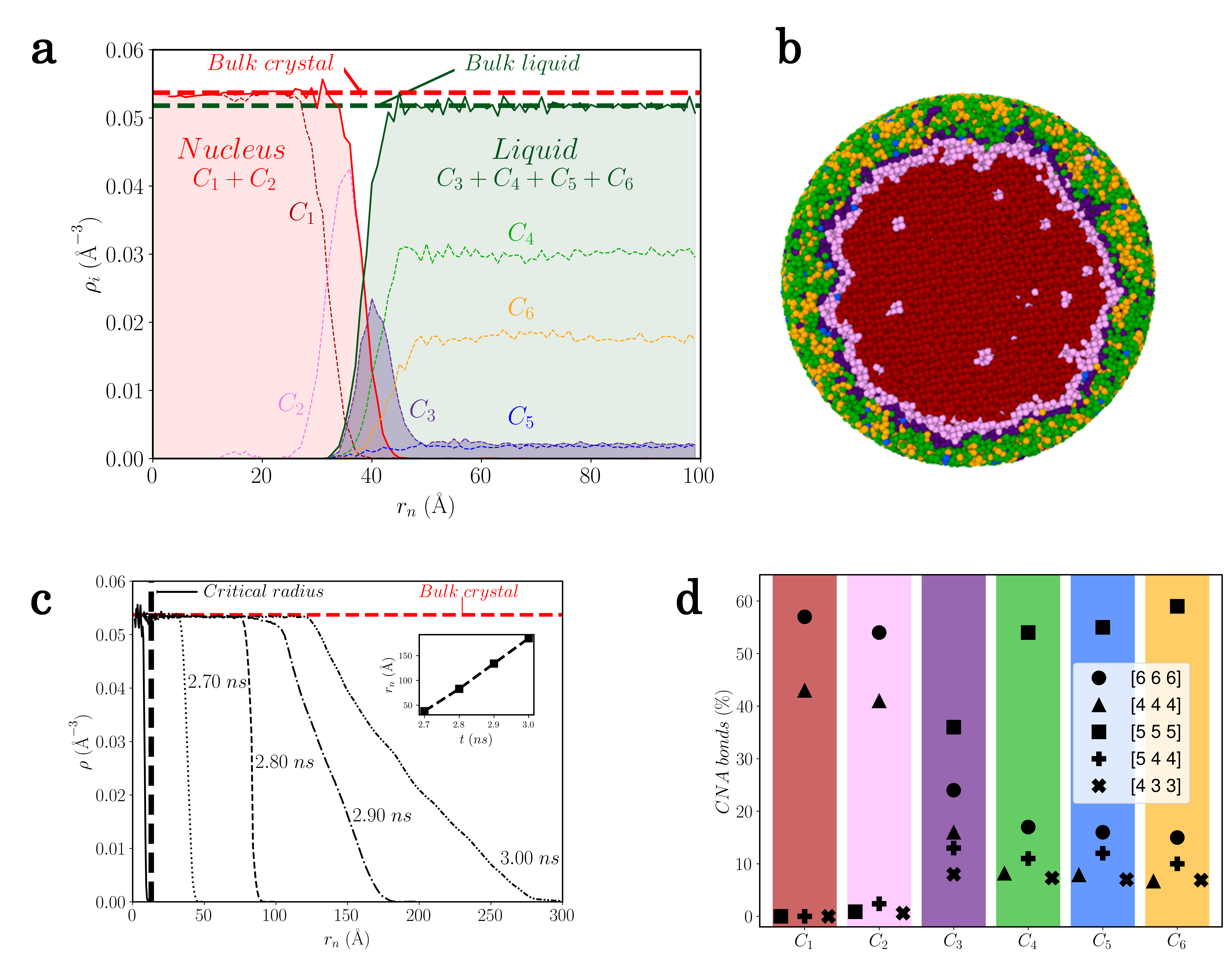}  
		\caption{\textbf{Translational and bond-orientational order parameters for Ta.} (a) Radial density profile of the largest nucleus during the growth at $2.7$ ns along the $T=1900$ K isotherm. The red and blue dashed horizontal lines correspond respectively to the average bulk crystalline density and average bulk undercooled liquid without nucleation events, both being simulated at $T=1900$~K at ambient pressure (b) Corresponding slice of the nucleus through its centre and the surrounding liquid where atoms have been coloured according to the cluster they belong to (see Fig. \ref{fig:Fig1-method}(d)). (c) Total radial density profile of the largest nucleus during growth at times corresponding to Fig. \ref{fig:Fig2-observation} before solidification. Inset: time evolution of the radius of the nucleus. (d) Bond-orientational order in terms of bonded pairs of the Common-Neighbor Analysis \cite{Faken1994} for each cluster of the model.}
		\label{fig:Fig3-analysis}
	\end{figure}
	
	The nucleation process is characterized at least by two order parameters, the translational order (TO) and the crystalline ordering called hereafter the bond orientational order (BOO). A dedicated representation of the TO is the number density. It is primarily applied to the embryos and the nuclei at different stage of the growth, through the radial partial atomic density profiles $\rho_i(r) = N_i(r)/\frac{4\pi}{3} [(r+\Delta r)^3-r^3]$ as a function of distance $r$ of the estimated centre of the nucleus, $N_i(r)$ being the number of atoms belonging to cluster $C_i$ in a spherical shell of radius $r$ and thickness $\Delta r=1$ \AA. Considering the nucleation process of Ta as an illustration, Fig. \ref{fig:Fig3-analysis}(a) depicts the density profiles $\rho_i(r)$ for all $6$ clusters for the largest nucleus shown in Fig. \ref{fig:Fig2-observation}(a) and its surrounding liquid at time $2.7$~ns. The corresponding slice of the nucleus through its centre is drawn in Fig. \ref{fig:Fig3-analysis}(b). Thus, the nucleus is defined by atoms belonging to clusters $C_1$ and $C_2$ as described above, atoms of $C_1$ forming the centre of the nucleus, while atoms of $C_2$ being mainly located at its border, as can be easily confirmed visually. It should be noted that atoms of cluster $C_3$ are mainly located at the boundary of the nucleus, but they cannot be considered as being part of it, as they are also present in the entire box. From the total density profile of the nucleus $\rho_N(r) = \rho_1(r)+\rho_2(r)$, it can be seen clearly that the density of nucleus has already reached at this stage the one of the bulk crystal at the same temperature. Defining the remaining clusters ($C_3$ to $C_6$) as belonging to the liquid yields to a total density profile $\rho_L(r) = \sum_{i=3}^6\rho_i(r)$ showing that even in the vicinity of the nucleus the liquid is negligibly influenced by its presence, keeping the density of the bulk undercooled liquid. 
	
	Fig. \ref{fig:Fig3-analysis}(c) shows the evolution of the density profile $\rho_N(r)$ at different times of the growing process. The average radius $r_N$ of the nucleus is taken as the value of $r$ at half-maximum of $\rho_N(r)$ and its evolution with time is shown in the inset, displaying a linear behaviour in agreement with CNT \cite{Sosso2016}. Whatever the size of the nuclei, the density of the inner part is close to the bulk crystal. More importantly, this is all the more true for all the embryos below the critical size up to a single local structure of type $C_1$ or $C_2$ corresponding to the minimal size of about $65$ atomic structures identified by the TDA-GMM given the chosen cutoff radius (see Supplementary Information). This feature appears to be general as similar results are found for Al and Mg as shown in the Supplementary Information. 
	
	The BOO of each cluster is identified through the Common Neighbour Analysis (CNA) \cite{Faken1994}, chosen as a well-known and robust tool. The CNA signature \cite{Jakse2006} given in Fig. \ref{fig:Fig3-analysis}(d) reveals that structures from clusters $C_1$ and $C_2$ possess respectively a perfect and slightly distorted bcc crystalline ordering confirming the above analysis of nucleation and growth in terms of topological descriptors. Structures from clusters $C_4$, $C_5$ and $C_6$ display various high degrees of five-fold symmetry (FFS) characteristic of the liquid state together with a small but non negligible degree of bcc ordering, while structures from cluster $C_3$ retains both FFS and bcc order in similar proportions. Such a BOO of the four clusters associated to the liquid agrees well with \textit{ab initio} molecular dynamics simulations \cite{Jakse2004} and was interpreted as compatible with the A15 crystalline phase. This analysis in terms of CNA highlights and confirms that the TDA-GMM unsupervised learning approach is a powerful method to capture the structural picture in its finest details. 
	
	\begin{figure}[tb!]
		\includegraphics[scale=0.25]{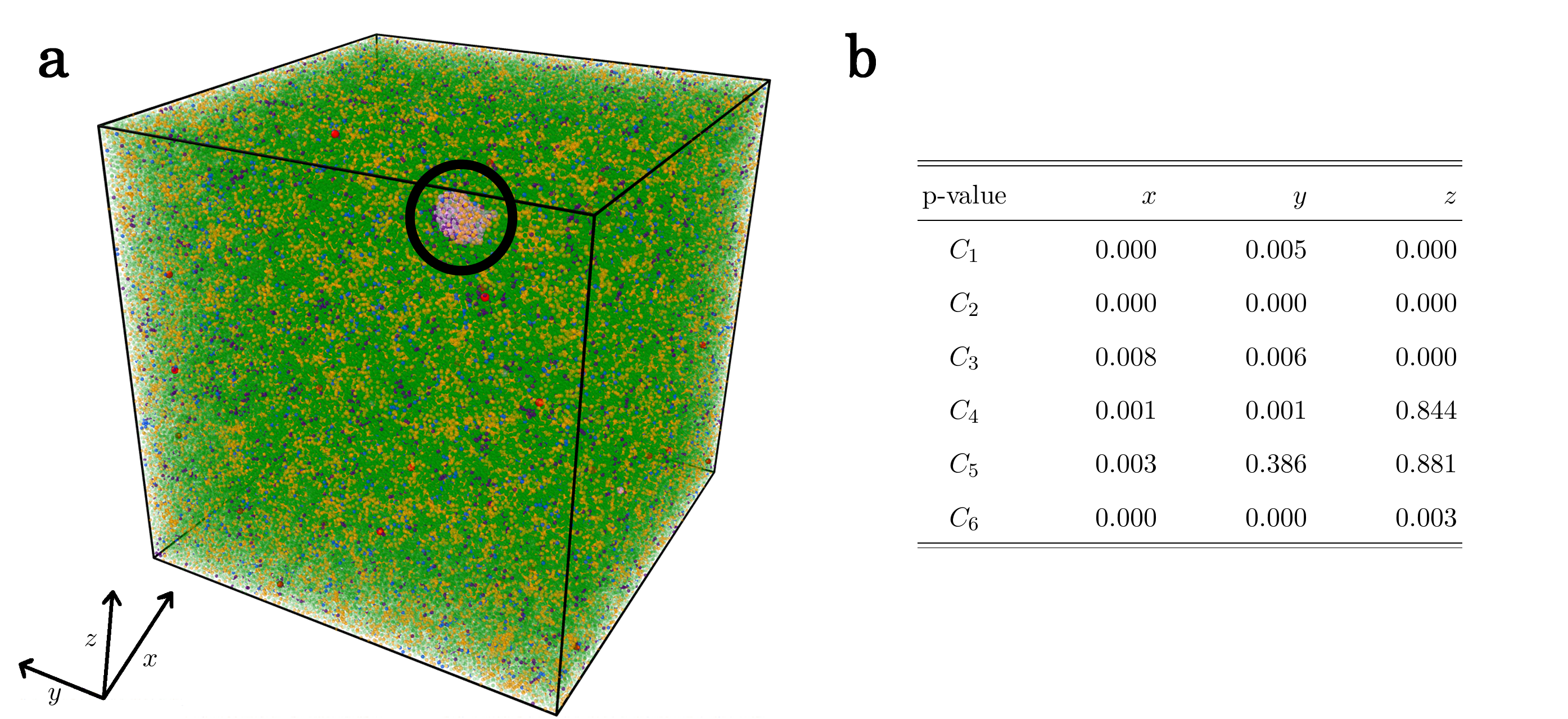}
		\caption{\textbf{Early nucleation stage for Ta.} (a) Snapshot of the simulation with $1$ million atoms along isotherm $T=1900 K$ showing the onset of nucleation when the first nucleus starts to grow (highlighted by the black circle). Atoms have been coloured according to the cluster they belong to (see Fig. \ref{fig:Fig1-method}(d)). (b) p-values computed on the projection of atomic positions on each direction of the box from a Kolmogorov-Smirnov test against the uniform distribution.}
		\label{fig:Fig4-onset-heterogeneity}
	\end{figure}
	
	The peculiar spatial distribution of structure of type $C_3$ shown in Fig.~\ref{fig:Fig3-analysis}(a) deserves further attention. Firstly, its location at the boundaries of the nuclei is consistent with the mixed bcc and FFS orderings. This is however seen as an effect of the TDA-GMM procedure that picks up structures covering a part of the nucleus and of the liquid in the configuration used for the training. More interestingly, its presence in the whole simulation box indicates that in the undercooled liquid, some regions with higher bcc ordering might develop apart from the vicinity of the growing nuclei. Fig.~\ref{fig:Fig4-onset-heterogeneity}(a) shows a snapshot of the simulation at the onset of nucleation when the first nucleus starts to grow, all atoms being coloured according to the cluster they belong to. 
	Fig.~\ref{fig:Fig4-onset-heterogeneity}(b) depicts a table of the concurrent p-values obtained, for each cluster, on the projection of atomic positions on the 3 directions of space, from a Kolmogorov-Smirnov test \cite{Kol33} against the uniform distribution. For a level 0.01, the test is always rejected in at least one direction, which proves that the distribution of the clusters in the box is not uniform, i.e. their heterogeneity. Focusing on atoms of type $C_4$ (green) and $C_6$ (yellow), which represent more than 90 \% of the atoms at this stage (58\% for $C_4$ and 34\% for $C_6$), it clearly shows that the undercooled liquid embodies structural heterogeneities with varying degree of FFS. Moreover, higher bcc ordering characterized by structures of type $C_3$ appears in localized regions of lower FFS (green) from which, in most of the cases, embryos formed by structures from clusters $C_1$ and $C_2$ emerge. The same conclusion of structural heterogeneity is obtained for Mg and Al, with particularly low p-values for Mg (see Supplementary Information).
	
	The question whether the onset of nucleation is initiated primarily by translational or by orientational ordering is still open, and was debated during the last decade with a controversy essentially centred on the hard sphere and associated colloidal systems \cite{Berryman2016,Russo2016}. For Ta, the small emerging embryos at the onset of nucleation, corresponding to one structure of 55 to 70 atoms belonging to $C_1$ or $C_2$ with bcc crystalline BOO, show bond lengths of their bcc lattice close to the density of the bulk crystal at $T=1900$~K, a feature that also holds for the other metals investigated here. This provides evidence for the size of embryos that can be detected here: translational and bond-orientational orders appear simultaneously and rule out the scenario in which homogeneous nucleation is driven by BOO first \cite{Russo2012} for metallic systems. This view is consistent with the fact that, unlike hard spheres, metallic systems with strong bonding are more energy driven rather than entropy driven systems.
	
	All these features allow us to propose a nucleation pathway for the metals considered here. For Ta, our findings show a single step process with an onset of homogeneous nucleation taking place in low FFS domains of the heterogeneous liquid, where emerging bcc embryos have simultaneously the density of the bulk solid. After reaching the critical size, the nuclei grows in a rather globular shape with a bcc structure with a small amount of defects and a diffuse interface with decreasing bcc ordering. During the growth the surrounding liquid keeps the bulk liquid density. A similar one step nucleation pathway also holds for Al in which embryos emerge from the low FFS regions directly with the fcc bond ordering. The growing nuclei have here a more patchy morphology and a significant amount of fcc stacking faults. For Mg, a two steps process is identified as can be seen in the Supplementary Information: an onset of nucleation showing embryos having mainly a bcc ordering followed by growth of nuclei with a mixed fcc/hcp structure and some bcc ordering at the surface of the nuclei. In this case, the scenario is more akin to the Lennard-Jones case \cite{ten1995,tenWolde1996} following the Landau Theory in which the bcc precursor is favoured in the early stages of crystal nucleation \cite{Alexander1974} as well as the Ostwald step rule \cite{Ostwald1897} for which the primary crystal phase nucleating from the liquid is not necessarily the thermodynamic stable one. 
	
	The present unsupervised learning approach was shown to be a powerful tool to unravel the atomic scale mechanisms of crystal nucleation in monatomic metals. It allowed us  to reveal general aspects in the homogeneous nucleation process as well as specificities depending on the metallic element under consideration. Our results are in line with the emerging idea that heterogeneities which exist in the undercooled liquid \cite{Russo2016} play the foremost role in the onset of nucleation. For all metals, nucleation have been found to start in low FFS regions, which is consistent with Frank's argument \cite{Frank1952}, with translational and orientational ordering taking place simultaneously in emerging embryos. Moreover, embryos as well as nuclei during the growth possess the bulk crystal density driven by the metallic bond length while the surrounding liquid keeps the bulk liquid density in accordance with the classical nucleation theory \cite{Sosso2016}. However, our analysis reveals also some aspects beyond the CNT, such as nuclei having a diffuse interface with the surrounding liquid and metals possessing their own nucleation pathways, involving \textit{e.g.} for Mg a two step mechanism \cite{Ostwald1897}. The complexity and richness found here for metals and in other systems \cite{Berryman2016,Russo2016,tenWolde1996} underline the future challenges in stepping forward in our theoretical understanding beyond the CNT. This promising methodology more generally opens the door to a deeper and autonomous investigation of atomic level mechanisms in materials science. 
	
	\section*{Methods}
	\textbf{Simulation method.} Molecular dynamics simulations were performed with the \textsc{lammps} code \cite{LAMMPS} in a fully periodic situation. Verlet’s algorithm in the velocity form for the numerical integration of the phase space trajectory was used with a time step of $2$ fs for Ta with a number of atoms $N=10^7$ ($10^6$ for the training) and $1$ fs for Al and Mg with $N=10^6$. Interatomic interaction were taken in the in Embedded Atom Model form and chosen for their ability to reproduce the liquid and solid properties as described in the Supplementary Information. Control of the thermodynamic conditions was done with the Nosé-Hoover thermostat and barostat \cite{Allen2017} was used to maintain the ambient pressure whatever the temperature. The time-temperature transformation curves were first built for each metal following the procedure established recently \cite{Becker2020}. Along an isotherm located slightly above the TTT nose, $6$ configurations of interest were selected for the purpose of monitoring the crystal nucleation process. Before analysing the configurations, minimization of the energy by means of a conjugate gradient algorithm has been performed to bring the system in a local minimum of the potential energy surface to suppress the thermal noise \cite{Stillinger1982}.
	
	\textbf{Persistent homological descriptors' space (TDA).} The unsupervised learning in the MD configurations is performed in terms of the local atomic environment of each atom (called the local structure) within a cut-off radius defined as the second minimum of the pair-correlation function $g(r)$ in the liquid, as described in Fig.~S1 of Supplementary Information. The use of two atomic neighbour shells to represent the local environment was shown to optimize the local structural information of descriptors at the expense of a loss of the spatial resolution \cite{Dellago2008}. In Persistent Homology \cite{Motta2018,Carriere2015}, components of homological dimensions $H_{0}$, $H_{1}$ and $H_{2}$ are then used in the form of a topological vector of dimension $n_{PH}$ to represent each local structure. Its components are calculated from the Persistent Diagrams (PD) representing the birth and death characteristics of each topological component, as shown in Fig.~S2 of the Supplementary Information. More precisely,
	 for each pair of points $(x,y)$ in a PD, $D$, the values of the topological vector components are calculated, except for the infinite point, for a fixed level of homology \cite{Carriere2015} by
	\begin{equation}
		m_D(x,y) = \min \{\|x-y\|_\infty, d_\Delta(x), d_\Delta(y)\},
	\end{equation}
	where $d_\Delta(\cdot)$ denotes the $\ell^\infty$ distance to the diagonal. The number of $H_0$ is fixed by the number of neighbour atoms and the number of components of $H_1$ and $H_2$ is inferred from a subsampling approach as described in \cite{Fasy2014} to remove the noise. 
	
	\textbf{Clustering using a Gaussian mixture Model (GMM).} In order to build a training set for the learning, a sampling of $5$ $000$ to $7$ $000$ structures, that covers the entire simulation box by means of their central particles at least separated by two times a cut-off radius are extracted from a million atoms configuration chosen during the nucleation.
	From the build topological descriptors' space as described above, a mixture of $M$ Gaussian distributions $(\phi(\;\cdot\;;\boldsymbol{\mu}_{m},\Sigma_{m}))_{1\leq m \leq M}$ of weights $(\alpha_{m})_{1\leq m \leq M}$ as
	\begin{equation}
	\sum_{m=1}^{M}\alpha_{m}\phi(\;\cdot\;;\boldsymbol{\mu}_{m},\Sigma_{m}),
	\end{equation}
	where $\boldsymbol{\mu}_{m}$ is the position of the mean and $\Sigma_{m}$ the covariance matrix of the $m$th Gaussian distribution. The number of Gaussian components is set using the ICL criterion \cite{ICL} and full covariance matrices with $3$ $000$ K-means initializations are used to construct a model for applications on configurations along the nucleation process.
	
	\section*{Acknowledgments}
	We acknowledge the CINES and IDRIS under Project No. INP2227/72914, as well as CIMENT/GRICAD for computational resources. This work was performed within the
	framework of the Centre of Excellence of Multifunctional Architectured Materials “CEMAM” ANR-10-LABX-44-01 funded by the “Investments for the Future” Program. This
	work has been partially supported byMIAI@Grenoble Alpes (ANR-19-P3IA-0003). Fruitful discussions within the French collaborative network in high-temperature thermodynamics
	GDR CNRS 3584 (TherMatHT) are also acknowledged.

\renewcommand{\thefigure}{S\arabic{figure}}
\renewcommand{\thetable}{S\Roman{table}}

\begin{center}
	{\Large Supplementary Information File}
\end{center}
\maketitle

Additional information are presented here to support data and figures of the main text.

\section{Supplementary informations on the methodology}
\subsection{Molecular dynamics simulations}

Table \ref{Tab:ST-Parameters} presents some characteristic properties related to each system described by potentials that are used (Ta \cite{Zhong2014}, Al \cite{Mendelev2008} and Mg \cite{Wilson2016}) in the present classical molecular dynamics simulations. Namely: the melting temperature $T_{m}$; the glass transition temperature $T_{g}$ which were extracted from the quenching of the liquid at ambient pressure with a cooling rate $Q$ up to the amorphous state; the isotherm $T_{\textrm{iso}}$ along which the nucleation process was studied; the ratios $T_{rg}=T_{g}/T_{m}$ and $\Delta{T}=(T_{m}-T)/T_{m}$; the estimated critical size of the nuclei $n_{c}$; and the critical cooling rate $Q_{c}$, up to which the crystallization can be avoided, inferred from the nose of the TTT curves. 

The procedure to compute the TTT curves follows the one from our previous work on Zr \cite{Becker2020}.

\begin{table}[H]
	\centering
	\begin{tabular}{crrrrrrrr}
		\hline
		\hline
		& \makebox[1.5cm][r]{$T_{m}$ (K)} & \makebox[1.5cm][r]{$T_{g}$ (K)} & \makebox[1.7cm][r]{$T_{\textrm{iso}}$ (K)} & \makebox[1.5cm][r]{$T_{rg}$} & \makebox[1.5cm][r]{$\Delta{T}$} & \makebox[1.5cm][c]{$n_{c}$} & \makebox[1.5cm][r]{$Q$ (K/s)}& \makebox[2cm][r]{$Q_{c}$ (K/s)} \\
		\hline
		Ta & $3290^{a}$ & 1582 & 1900 & 0.48 & 0.42 & 140-150 & $10^{12}$ & $4.2\times10^{11}$ \\
		Al & $926^{b}$ & 291 & 650 & 0.31 & 0.30 & 400-800 & $10^{13}$ & $1.8\times10^{12}$ \\
		Mg & $918^{c}$  & 303 & 600 & 0.33 & 0.35 & 310-350 & $10^{12 }$& $3.2\times10^{11}$ \\
		\hline
	\end{tabular}
	\caption{Characteristic features of the classical molecular dynamics potentials as described in the text. Melting temperatures $T_{m}$ are taken from Refs. $^{a}$\cite{Zhong2014}; $^{b}$\cite{Mendelev2008}; $^{c}$\cite{Wilson2016}}
	\label{Tab:ST-Parameters}
\end{table}

\subsection{Definition of the local structures}

\begin{figure}[H]
	\centering
	\includegraphics[width=1\textwidth]{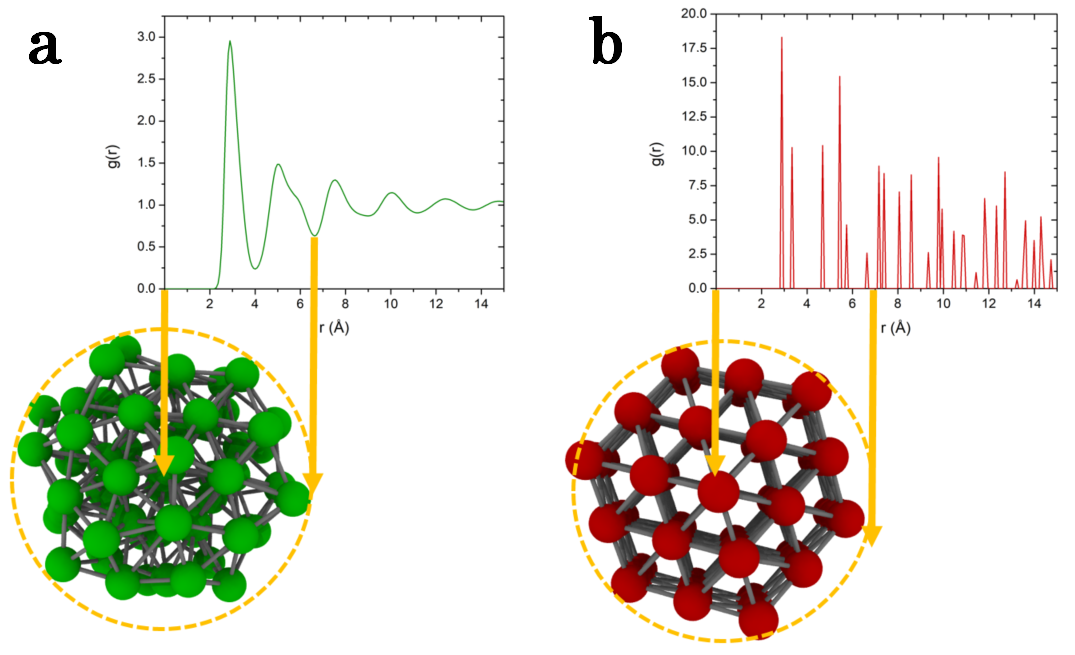}
	\caption{Cut-off radius for the clustering based on the $g(r)$ function with the second minimum leading to structures with two neighbors shells with the one associated in Ta to the preponderant liquid in green (a) and the one to bcc ordering in red (b).}
	\label{fig:SFig1-cutoff}
\end{figure}

Figure \ref{fig:SFig1-cutoff} shows the pair-correlation function $g(r)$ of the undercooled liquid and crystalline states of Ta at $T=1900$ K with the respective mean structure assigned to the clusters $C_{4}$ (preponderant liquid) and $C_{1}$ (preponderant bcc). A cut-off radius of $6.8$ $\AA$ was set to capture topological informations with the help of the Python package \texttt{gudhi} \cite{Maria2014} and \texttt{ripser.py} \cite{Tralie2018} up to the second neighbour shell. 

In the context of classical descriptors like the averaged bond-orientational order analysis it was shown \cite{Dellago2008} that information from the second neighbour shell increase the accuracy in the discrimination of local structures, but at the expense of a loss in the spatial resolution. This is also observed in the topological descriptors set up here with give rise to more $H_0$ and $H_1$ components as well $H_2$ components which appear only when considering more than just one neighbour shell. It should be pointed our that when increasing further the cut-off radius up to the third neighbour shell and beyond, the benefit gained in topological information is counterbalanced by a too large spatial extension leading to a loss of resolution in the Gaussian Mixture Model (GMM) clustering. This compromise between the accuracy and the spatial resolution bring us the optimal choice of the second neighbour shell to define local structures consistently with earlier findings \cite{Dellago2008}.

\begin{figure}[H]
	\centering
	\includegraphics[width=1\textwidth]{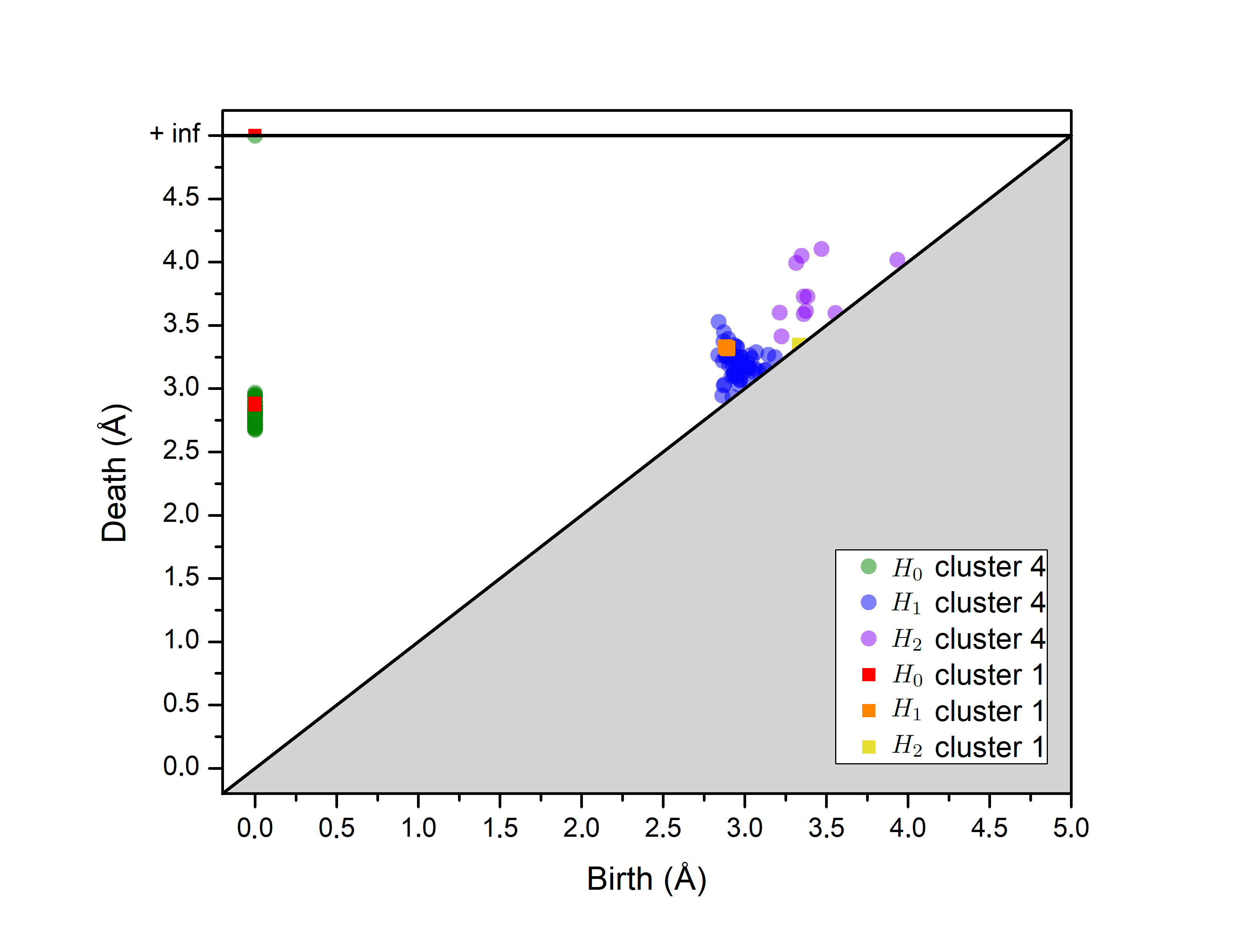}
	\caption{Persistence diagrams for the mean structures of $C_{1}$ and $C_{4}$ which represent respectively the bcc ordering and the preponderant liquid structure.}
	\label{fig:SFig2-PD}
\end{figure}

The local atomic structures were extracted with Python package \texttt{pyscal} \cite{Menon2019}. For comparison, the persistent homological information are depicted on the persistence diagram shown in Figure \ref{fig:SFig2-PD} for the two mean local structures assigned to $C_{4}$ and $C_{1}$. The differences can be seen here between a disordered liquid structure and a perfect periodic lattice where all the pairs (birth, death) are concurrent for each homological dimensions.

\subsection{Clustering using a Gaussian mixture Model (GMM)}

\begin{figure}[H]
	\centering
	\includegraphics[width=1\textwidth]{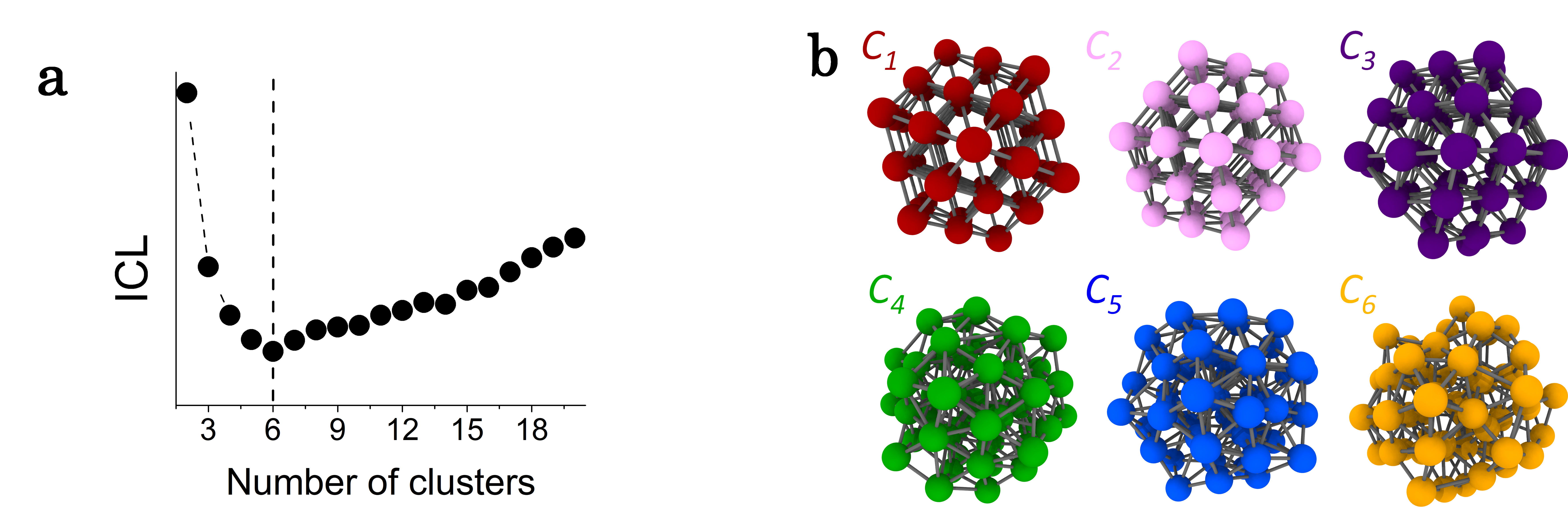}
	\caption{TDA-GMM clustering of Al. (a) Evolution of the Integrated Completed Likelihood (ICL) criterion as a function of number of clusters. (b) Independent local atomic structures within a cut-off-radius of $6.3$ $\AA$ form a train set represented in the descriptor space by $173$ PH components up to the second order.}
	\label{fig:Al_S1}
\end{figure}

\begin{figure}[H]
	\centering
	\includegraphics[width=1\textwidth]{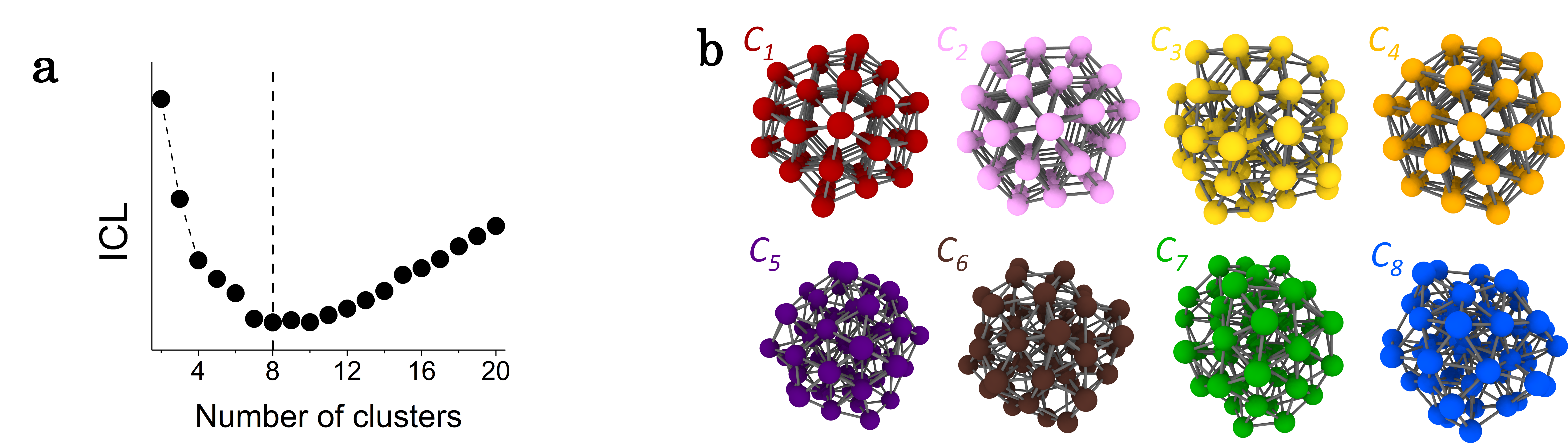}
	\caption{TDA-GMM clustering of Mg. (a) Evolution of the Integrated Completed Likelihood (ICL) criterion as a function of number of clusters. (b) Independent local atomic structures within a cut-off-radius of $6.9$ $\AA$ form a train set represented in the descriptor space by $199$ PH components up to the second order.}
	\label{fig:Mg_S1}
\end{figure}

Figures \ref{fig:Al_S1} and \ref{fig:Mg_S1} shows the clustering with the TDA-GMM method applied respectively to Al and Mg configuration during nucleation. The resulting Local atomic structures assigned to each cluster are shown. The number of clusters is determined using the ICL criterion. The clustering is performed with Python package \texttt{scikit-learn} \cite{Pedregosa2011}.

\section{Crystal nucleation}
\subsection{Identification and extraction of the structures during nucleation}	

Tables \ref{Tab:ST-ClustersTa}, \ref{Tab:ST-ClustersAl} and \ref{Tab:ST-ClustersMg} show respectively the evolution of proportion of the structures assigned to each clusters previously identified by the TDA-GMM method in Ta, Al and Mg. As can be seen in each case, clusters $C_{1}$ and $C_{2}$ follow a fast growth of their proportion along the nucleation process until they are the majority in the final bulk. Referring to Figure $1$ in the main text and the previous Figures \ref{fig:Al_S1} and \ref{fig:Mg_S1}, these two clusters are indeed represented by local atomic structures on a pure and distorted crystalline structures. In the case of Mg, one can notice that clusters $C_{3}$ and $C_{5}$ are also growing along $C_{1}$ and $C_{2}$. This is explained by the fact that part of the structures assigned to these two clusters shares bonds with the growing nuclei and carry partial crystalline structure owing to the use a cut-off corresponding to the second neighbour shell. 

\begin{table}[H]
	\centering
	\begin{tabular}{crrrrrr}
		\hline
		\hline
		Time (ns) & \makebox[2cm][r]{2.70} & \makebox[2cm][r]{2.80} & \makebox[2cm][r]{2.90} & \makebox[2cm][r]{2.96$^{(M)}$} & \makebox[2cm][r]{3.00} & \makebox[2cm][r]{3.60$^{(S)}$}\\
		\hline
		$C_{1}$ (\%) & 0.09 & 1.03 & 5.18 & 13.44 & 15.84 & 83.52 \\
		$C_{2}$ (\%) & 0.06 & 0.33 & 1.25 & 4.81 & 3.27 & 12.25 \\
		$C_{3}$ (\%) & 4.13 & 4.10 & 4.12 & 5.61 & 4.18 & 2.92 \\
		$C_{4}$ (\%) & 58.11 & 57.10 & 53.72 & 45.56 & 45.88 & 0.85 \\
		$C_{5}$ (\%) & 3.39 & 3.38 & 3.25 & 3.05 & 2.84 & 0.20 \\
		$C_{6}$ (\%) & 34.23 & 34.06 & 32.48 & 27.53 & 28.00 & 0.27 \\
		\hline
		\hline
	\end{tabular}
	\caption{Proportion of each cluster for Ta at different times during the nucleation process. Superscripts (M) and (S) correspond respectively to the configuration used to train the TDA-GMM model and the solidified configuration.}
	\label{Tab:ST-ClustersTa}
\end{table}	

\begin{table}[H]
	\centering
	\begin{tabular}{crrrrrr}
		\hline
		\hline
		Time (ps) & \makebox[2cm][r]{166} & \makebox[2cm][r]{170} & \makebox[2cm][r]{174} & \makebox[2cm][r]{175$^{(M)}$} & \makebox[2cm][r]{177} & \makebox[2cm][r]{240$^{(S)}$}\\
		\hline
		$C_{1}$ (\%) & 0.05 & 0.54 & 3.42 & 4.24 & 10.07 & 36.55 \\
		$C_{2}$ (\%) & 0.73 & 2.29 & 7.14 & 9.10 & 14.34 & 22.43 \\
		$C_{3}$ (\%) & 19.78 & 19.42 & 19.74 & 19.7 & 20.29 & 23.57 \\
		$C_{4}$ (\%) & 33.22 & 32.46 & 29.32 & 28.85 & 23.76 & 9.95 \\
		$C_{5}$ (\%) & 44.38 & 43.53 & 38.94 & 36.92 & 30.60 & 7.48 \\
		$C_{6}$ (\%) & 1.84 & 1.76 & 1.43 & 1.19 & 0.95 & 0.02 \\
		\hline
		\hline
	\end{tabular}	
	\caption{Proportion of each cluster for Al at different times of the nucleation process. Superscripts (M) and (S) correspond respectively to the configuration used to train the TDA-GMM model and the solidified configuration.}
	\label{Tab:ST-ClustersAl}
\end{table}

\begin{table}[H]
	\centering
	\begin{tabular}{crrrrrr}
		\hline
		\hline
		Time (ps) & \makebox[2cm][r]{940} & \makebox[2cm][r]{960} & \makebox[2cm][r]{980} & \makebox[2cm][r]{990$^{(M)}$} & \makebox[2cm][r]{1000} & \makebox[2cm][r]{1500$^{(S)}$}\\
		\hline
		$C_{1}$ (\%) & 0.01 & 0.13 & 0.74 & 3.82 & 4.60 & 20.59 \\
		$C_{2}$ (\%) & 0.11 & 0.34 & 1.32 & 3.64 & 5.28 & 14.58 \\
		$C_{3}$ (\%) & 0.31 & 1.13 & 3.00 & 3.98 & 6.39 & 12.25 \\
		$C_{4}$ (\%) & 0.20 & 1.37 & 4.38 & 3.64 & 2.38 & 1.13 \\
		$C_{5}$ (\%) & 4.14 & 5.03 & 6.68 & 8.56 & 10.56 & 18.30 \\
		$C_{6}$ (\%) & 23.91 & 23.15 & 21.59 & 20.78 & 20.68 & 19.32 \\
		$C_{7}$ (\%) & 36.64 & 35.30 & 32.12 & 28.56 & 25.30 & 5.16 \\
		$C_{8}$ (\%) & 34.68 & 33.56 & 30.17 & 27.01 & 24.81 & 8.68 \\
		\hline
	\end{tabular}	
	\caption{Proportion of each cluster for Mg at different times of the nucleation process. Superscripts (M) and (S) correspond respectively to the configuration used to train the TDA-GMM model and the solidified configuration.}
	\label{Tab:ST-ClustersMg}
\end{table}

\begin{figure}[H]
	\centering
	\includegraphics[width=1\textwidth]{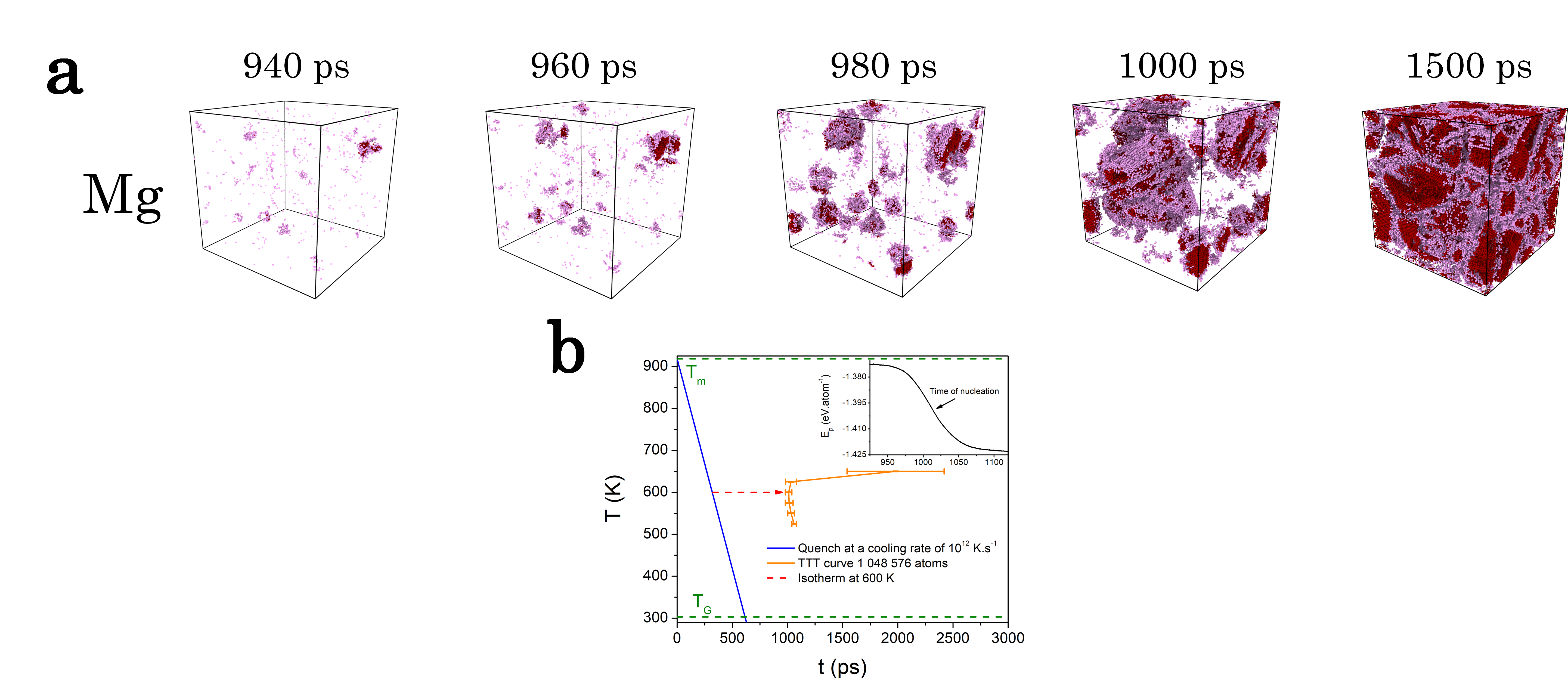}
	\caption{Homogeneous nucleation events in undercooled Mg during an isothermal process (a) at the nose of the TTT curve (b).}
	\label{fig:Mg_S3}
\end{figure}

Figure \ref{fig:Mg_S3} shows the evolution in Mg of the central particles assigned to $C_{1}$ and $C_{2}$ through the nucleation process along with the TTT curve.

\subsection{Translational and orientational orderings}

\begin{figure}[H]
	\centering
	\includegraphics[width=1\textwidth]{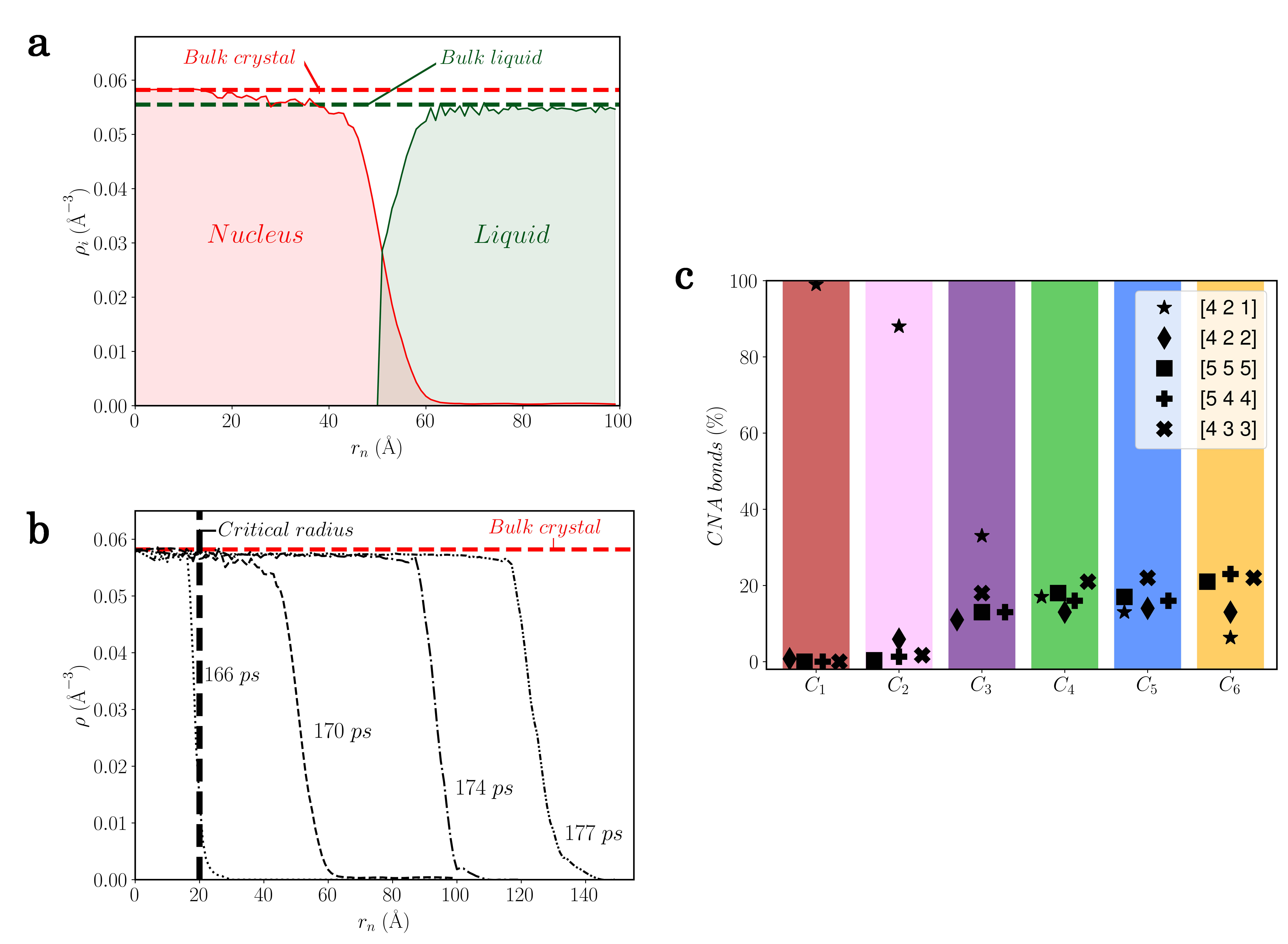}
	\caption{Typical translational (a) and bond-orientational (c) order parameters for Al. An analysis of the density profile at various times of the biggest growing nuclei (b) shows that the translational order is concurrent with the orientational order at the onset of nucleation.}
	\label{fig:Al_S2}
\end{figure}

\begin{figure}[H]
	\centering
	\includegraphics[width=1\textwidth]{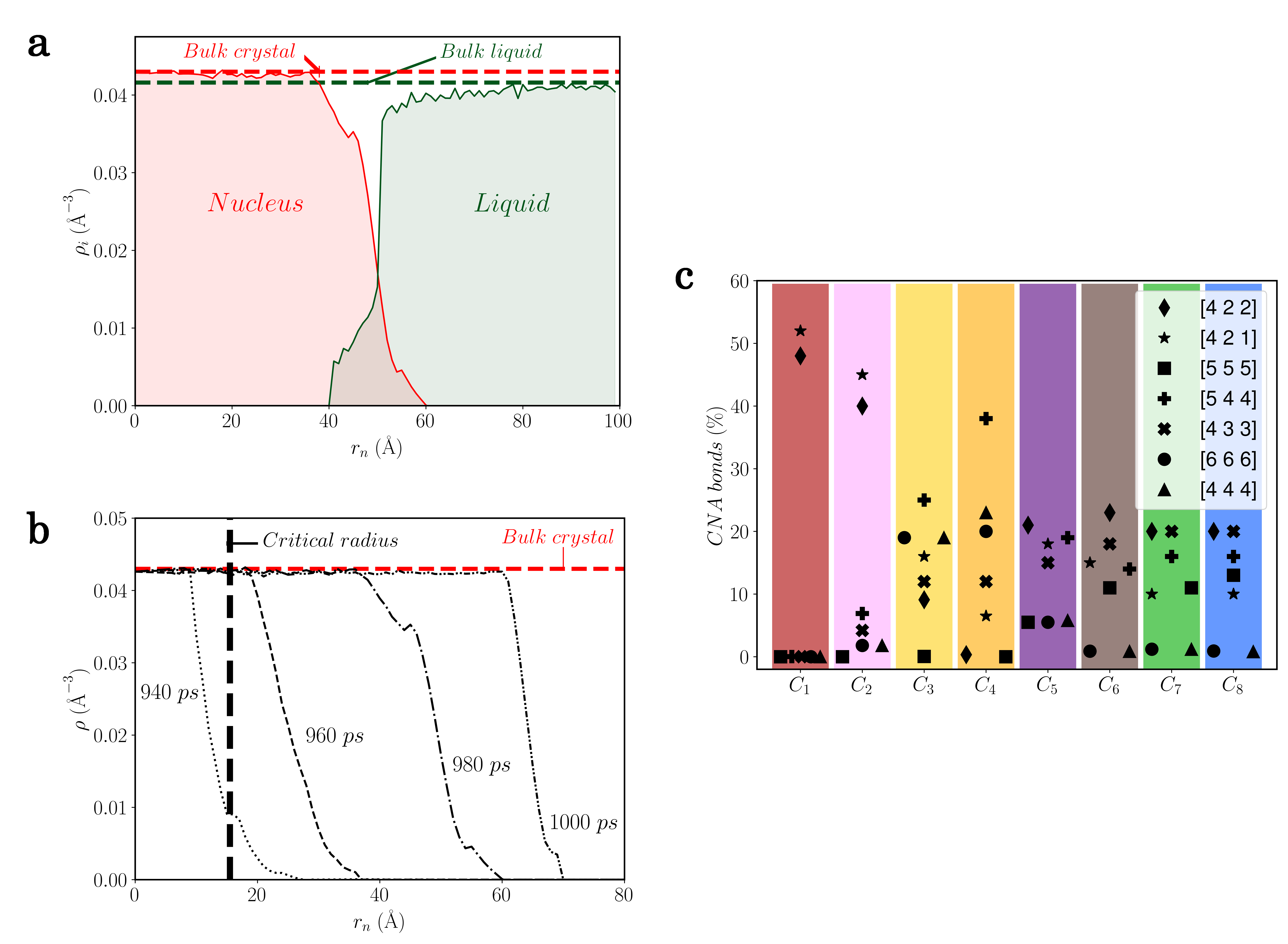}
	\caption{Typical translational (a) and bond-orientational (c) order parameters for Mg. An analysis of the density profile for various times of the biggest growing nuclei (b) shows that the translational order is concurrent with the orientational order at the onset of nucleation.}
	\label{fig:Mg_S2}
\end{figure}

Following the procedure described in the main text, a general behaviour for the translational and orientational orderings of Al and Mg is depicted on the Figures \ref{fig:Al_S2} and \ref{fig:Mg_S2}. All the nuclei are driven by a concurrent emergence of this two symmetries which correspond respectively to the density of the crystal bulk and the geometrical bonds related to the crystalline local structure.

\begin{thebibliography}{99}
\bibitem{warningCluster} The word 'cluster' is used for groups detected by the machine learning method throughout the text.
\end{thebibliography}
\end{document}